\documentclass[journal,12pt,draftclsnofoot,onecolumn]{IEEEtran}

\usepackage{amsfonts}
\usepackage{amsmath}

\usepackage{cite}      

\usepackage{graphicx}  

\usepackage{stfloats}  

\usepackage{tablefootnote}

\newcommand{\comment}[1]{}
\newcommand{\mycomment}[1]{}
\newcommand{\gmcomment}[1]{}
\newcommand{\obmcmntout}[1]{} %
\newcommand{\obmreuse}[1]{} %
\newcommand{\obmsubmittodo}[1]{} %
\newcommand{\psbfg}[1]{} %

\newcommand{\obmabrdgol}[1]{} %
\newcommand{\obmthg}[1]{} 

\newcommand{\figcomment}[1]{}

\begin{document}

\title{MUSE: A Methodology for Characterizing and Quantifying the Use of Spectrum}

\author{Nilesh~Khambekar,~\IEEEmembership{Member,~IEEE,}
        ~Chad~M.~Spooner,~\IEEEmembership{Senior~Member,~IEEE,}
        and~Vipin~Chaudhary,~\IEEEmembership{Member,~IEEE} 
\date{\copyright\ 2015 Nilesh Khambekar\  All Rights Reserved} 				
\thanks{Nilesh Khambekar and Vipin Chaudhary are with the Department of Computer Science and Engineering, University at Buffalo, SUNY, Buffalo, New York 14260-2000}
\thanks{Chad M. Spooner is with NorthWest Research Associates, Monterey, CA.}}


%


\maketitle

\begin{abstract}
Dynamic spectrum sharing paradigm is envisaged to meet the growing demand for the Radio Frequency (RF) spectrum. There exist several technical, regulatory, and business impediments for adopting the new paradigm. In this regard, we underscore the need of characterizing and quantifying the use of spectrum by each of the individual transmitters and receivers.

We propose MUSE, a methodology to characterize and quantify the use of spectrum in the space, time, and frequency dimensions. MUSE characterizes the use of spectrum by a transmitter at a point in terms of the RF power occupied by the transmitter. It characterizes the use of spectrum by a receiver at a point in terms of the constraints on the RF-power that can be occupied by any of the transmitters in the system in order to ensure successful reception. It divides the spectrum-space into discrete unit-spectrum-spaces and quantifies the spectrum used by the individual transceivers in the discretized spectrum space. 

We characterize the performance of the spectrum management functions in the discretized spectrum-space and illustrate maximizing the use of spectrum. In order to address the challenges for the dynamic spectrum sharing paradigm, we emphasize on articulating, defining, and enforcing the spectrum-access rights in the discretized spectrum-space.
\end{abstract}



%
\IEEEpeerreviewmaketitle


\setlength{\textfloatsep}{10pt}
\setlength{\intextsep}{0mm}


\section{Introduction}
Traditionally, the radio frequency (RF) spectrum has been statically and exclusively allocated.  This static spectrum allocation paradigm results into an inefficient usage of the spectrum in time, space, and frequency dimensions \cite{fcc_sptf, ssc_erpek}. In order to meet the growing demand for the new and high bandwidth wireless services, the spectrum needs to be dynamically shared by multiple wireless service providers \cite{fcc_nprm_2003, pcast, dod_ems}. 

The dynamic spectrum sharing paradigm presents new challenges on technical, regulatory, and business fronts. For effective spectrum sharing, non-harmful interference needs to be ensured among multiple heterogeneous RF-systems under the dynamic RF-environment conditions. With the static and exclusive spectrum allocation paradigm, the spectrum-access parameters for a service are chosen so as to mitigate potential interference and ensure minimum performance under the worst-case conditions. Defining spectrum sharing constraints to ensure \textit{minimum} performance under the \textit{worst-case} propagation conditions severely limits the opportunities to exploit the underutilized spectrum \cite{smm_thesis, berk_wsc, osa_feasib, oms2_sca}. We need the ability to define and enforce spectrum-access constraints that can maximize the availability and exploitation of the underutilized spectrum under dynamic RF environment conditions. In this regard, the aggregate interference effects, dynamic propagation conditions, and software defined capabilities bring in complexity to the regulation of dynamic spectrum-access. Furthermore, from a business perspective, it is also important to be able to flexibly and efficiently trade the spectrum in addition to solving the technical and regulatory issues.

In order to address the challenges for the adoption of the new paradigm, we investigate what constitutes the use of spectrum and emphasize the need to characterize the use of spectrum by each of the transmitters and receivers in the space, time, and frequency dimensions. We highlight the lack of the ability to quantify the performance of recovery and exploitation of the underutilized spectrum. We propose a Methodology to characterize and quantify the USE of spectrum in the space, time, and frequency dimensions (MUSE). MUSE is \textit{independent}\footnote{The system model considers a generic collection of transceivers. Thus, MUSE can be also be applied under traditional scenarios without spectrum sharing.} of the spectrum sharing models and can scale across various simple to advanced spectrum sharing use-cases.  By characterizing the use of spectrum, MUSE facilitates articulating the spectrum-access rights in terms of the use of spectrum. We argue that this ability is essential to address several technical, regulatory, and business difficulties. Furthermore, by characterizing the use of spectrum, MUSE enables characterization of the spectrum management functions in the space, time, and frequency dimensions. This ability helps us to optimize the performance of spectrum management functions in order to maximize the use of spectrum. 

The rest of this paper is organized as follows. In Section II, we underscore the need for a methodology to characterize and quantify the use of spectrum and the performance of spectrum management functions. In Section III, we present the mechanisms for characterizing and quantifying the use of spectrum in the space, time, and frequency dimensions using MUSE. In Section IV, we explain the methodology with a few examples and discuss the impact of the key factors while applying MUSE. In Section V, we describe how MUSE facilitates analysis, estimation, and optimization of the spectrum consumed by the transceivers and enables maximizing the use of spectrum. In Section VI,  we illustrate the benefits of MUSE for operations, regulations, and commerce of the spectrum. Finally, in Section VII, we draw conclusions and outline further research avenues.

\section{Motivation}

The dynamic spectrum sharing approaches have been evolving since the past decade \cite{dsa_survey, akyildiz_survey1, winn_survey}. Depending on the degree of sharing, the various spectrum sharing approaches fall into exclusive spectrum use, static spectrum sharing, dynamic spectrum sharing, and pure spectrum sharing categories \cite{winn_survey}. Dynamic spectrum sharing differs from pure spectrum sharing in the sense that under pure spectrum sharing all services have equal spectrum-access priority. Zhao \textit{et. al.} classified spectrum sharing approaches into open sharing model, dynamic exclusive use model, and hierarchical access model \cite{dsa_survey}. The hierarchical access model could be further categorized into spectrum underlay model, non-prioritized spectrum overlay model, and prioritized spectrum overlay model. Spectrum underlay model imposes tight constraints on secondary spectrum-access in order to protect the spectrum-access rights of the incumbents. Under non-prioritized spectrum overlay model, a secondary spectrum-access is granted on a first come, first served basis while ensuring non-harmful interference to the receivers of the incumbent services. Under prioritized spectrum overlay model, certain services are assigned priority access privileges and the secondary access by these services is protected. Other non-prioritized secondary spectrum accesses are required to vacate if a priority user wishes to access spectrum. The proposed 3.5 GHz Citizens Broadband Radio Service (CBRS) \cite{fcc_cbrs} is an example of prioritized spectrum overlay model.

In terms of articulating the spectrum access rights, the spectrum sharing mechanisms primarily resort to statically or dynamically defining a spatio-temporal boundary along with a \textit{fixed} set of constraints. In this regard, the case study of dynamic spectrum sharing in UHF bands has brought out several technical, regulatory, and business difficulties.

In Nov. 2008, Federal Communications Commission (FCC) released a Notice of Proposed Rule Making (NPRM) to allow the unlicensed radios to operate in the TV bands without causing harmful interference to the incumbent services \cite{fcc_nprm_2008}. The Opportunistic Spectrum Access (OSA) of the unused UHF bands received a wide commercial interest for several potential wireless services; However, the performance estimation studies of OSA have revealed that the amount of the \textit{implied} available spectrum is \textit{very limited} to meet the increasing demand for RF spectrum \cite{smm_thesis, berk_wsc, osa_feasib}. Moreover, the secondary users cannot ensure desired quality of service necessary for the business cases due to the \textit{secondary} rights for accessing the spectrum. On the other hand, incumbents do not have any incentive for sharing the spectrum. Furthermore, the secondary access to the spectrum is very hard to regulate. Considering interference aggregation effects, dynamic nature of propagation conditions, and dynamic spectrum-access scenarios, the primary owners of the spectrum need a way to confirm that their receivers are not subjected to harmful interference and the service experience is not degraded. This requires the ability to reliably estimate the interference margin at the receivers and accordingly infer the maximum transmit-power at the secondary transmitter positions. Furthermore, the behavior of software defined radio devices could be altered with software changes and thus the service is exposed to attacks from the secondary users of the spectrum. In order to ensure protection of the spectrum rights, the spectrum-access constraints need to be \textit{enforceable}. 

We observe that the decisions for exercising spectrum-access in case of OSA are based on detection of primary transmitter signal using a certain specified radio sensitivity. In this case, the decision for spectrum-access is binary in nature. This gives rise to `\textit{not enough spectrum for secondary usage}' if the policy for shared spectrum-access is conservative and `\textit{no guarantee for ensuring service quality}' if the shared spectrum-access policy is aggressive. The binary nature of the spectrum-access decision cannot protect the spectrum rights of incumbents and requires the spectrum-access policy to be increasingly conservative to guard against interference aggregation. Therefore, when multiple secondary transmitters exercise spectrum-access, we need to \textit{quantitatively articulate the spectrum-access rights}. This helps maximizing a spectrum-access opportunity without causing harmful interference. If technical and regulatory problems are solved, more and more incumbents will have an incentive to share the spatially, temporally, and spectrally unexploited spectrum. 

\begin{figure*}[htbp!]
\centering
{\includegraphics [width=0.92\textwidth, angle=0] {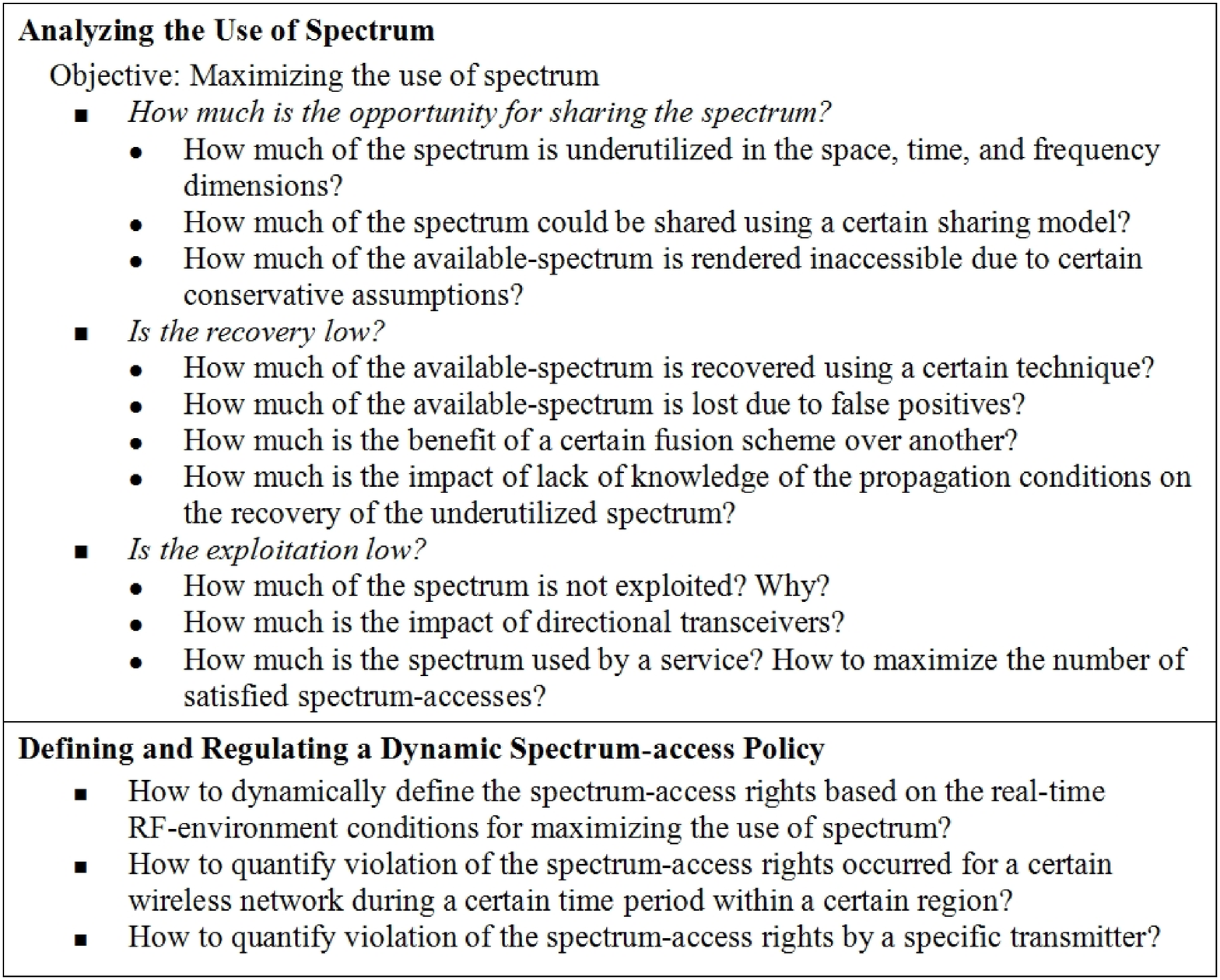}}
\setlength{\abovecaptionskip}{-9pt}
\caption{Example questions in case of optimizing a typical dynamic spectrum sharing scenario. The questions shade light on the various quantitative decisions involved with regards to spectrum sharing and spectrum management. The question-map emphasizes on the need for a methodology to characterize and quantify the use of spectrum in order to effectively manage the use of spectrum.}
\label{fig:qdsa_qneed}
\end{figure*}

Figure~\ref{fig:qdsa_qneed} illustrates the need for a methodology to characterize and quantify the use of spectrum under dynamic spectrum sharing paradigm with the aid of a question-map. The question-map enumerates the quantitative decisions involved in the process of investigating the weaknesses of a spectrum sharing mechanism, comparing various algorithms and architectures for recovery and exploitation of the spectrum, and optimizing the spectrum sharing opportunities. 

Traditionally the performance of spectrum recovery is measured in terms of the throughput for the secondary users and outage probability \cite{mishra_coop_sensing, ganesan, Visotsky}. The performance of detection of spectrum holes is also captured in terms of probability of missed detection and false positives \cite{ganesan_spthole, raman, tuando}. However, this characterization of the performance is in the context of spectrum sharing constraints defined by a certain spectrum sharing model or in terms of system-level objectives. In order to maximize the use of spectrum, we need a methodology that can characterize the performance of the recovery and exploitation of the underutilized spectrum in the space, time, and frequency dimensions. 

The existing methodologies to define the use of spectrum and quantify its efficiency are based on the static spectrum assignment paradigm and are not suitable for the dynamic spectrum sharing paradigm. ITU defined \textit{spectrum utilization factor} as product of the frequency bandwidth, geometric space, and the time denied to other potential users \cite{itumetrics}. However, spectrum utilization factor does not represent \textit{actual usage}. For example, if a licensed user does not perform any transmissions, the spectrum is still considered to be \textit{used}. It also cannot quantify the use of spectrum under spatial overlap of wireless services. The IEEE 1900.5.2 draft standard captures spectrum usage in terms of transceiver-model parameters and applies standard methods for ensuring compatibility between the spectrum sharing networks \cite{ieee_compat}. Thus, the approach helps to ensure compatibility; however, it cannot \textit{characterize and quantify} the use of spectrum and the performance of spectrum management functions.

Finally, from a business perspective, the ability to qualitatively and quantitatively interpret a spectrum sharing opportunity in a certain frequency band within a geographical region of interest is essential in order to evaluate its business potential. With the change in paradigm, businesses need the ability to control the use of spectrum at a fine granularity in order to maximize fine granular spectrum-reuse opportunities. With spectrum as a quantified resource perspective, the spectrum trade conversation could be on the following lines: \textit{``I have `x' units of spectrum right now, I have given `y' units of spectrum to somebody and have `z' units of spare spectrum which I can share".} Also, the quantification of the use of spectrum would provide insight into the business implications of a dynamically identified spectrum-access opportunity in terms of the service quality, range, and user experience. 

\section{MUSE: Characterizing and Quantifying the Use of Spectrum}
In order to define a methodology that enables us to characterize the use of spectrum in the space, time, and frequency dimensions, we first look into what constitutes the use of spectrum. 

\subsection{How is Spectrum Consumed?}
Traditionally, we assume that spectrum is consumed by the transmitters; however, the spectrum is \textit{also} consumed by the receivers by constraining the RF-power from the other transmitters. We note that for guaranteeing successful reception, protection is traditionally accomplished in terms of the guard-bands, separation distances, and constraints on the operational hours. Thus, the presence of receivers enforces limits on the interference-power in the space, time, and frequency dimensions. When the access to spectrum is exclusive in the space, time, and frequency dimensions, the spectrum consumed by the receivers need not be separately considered \cite{itumetrics}.
%
\subsection{System Model}
We consider a generic system with multiple heterogeneous \textit{spatially-overlapping}\footnote{Without allowing spatial-overlap of wireless services, spectrum sharing may lead to spatial fragmentation of coverage for a wireless service. Furthermore, as discussed in the previous section, imposing a spatial boundary on spectrum sharing leads to suboptimal spectrum sharing.}wireless services sharing the RF-spectrum.  We define a \textit{RF-link} represents \textit{zero}\footnote{This is to include the use of spectrum by the receiver-only systems; for example, radio astronomy telescopes.} or one transmitter and one or more receivers exercising spectrum-access. A \textit{RF-network} represents an aggregate of RF-links. We refer to the aggregate of RF-networks sharing a spectrum space in the time, space, and frequency dimensions within a geographical region of interest as a \text{RF-system}. We consider that a multiple RF-systems are sharing the spectrum in the time, space, and frequency dimensions within the geographical region of interest. 

We seek to capture spectrum-access at the lowest granularity. In this regards, RF-link represents the lowest granularity of spectrum-access.

Under the system model, we consider that the transceivers optionally employ directional transmission and reception in order to minimize interference. A receiver can withstand a certain interference when the received Signal to Interference and Noise Ratio (SINR) is greater than a receiver-specific threshold\footnote{The threshold, $\beta$, represents the quality of a receiver and incorporates receiver-noise and other receiver technology imperfections. Thus, $\beta$ models the receiver-performance under the proposed methodology.}, $\beta$. 

Let $P_{MAX}$ represent the maximum permissible power at any point and $P_{MIN}$ represent the minimum power at any point in the system. $P_{MIN}$ could be chosen to be an arbitrary low value below the thermal noise floor. The difference between the maximum and the minimum spectrum consumption at a point represents the maximum spectrum consumption, $P_{CMAX}$, at a point and it is given by
\begin{equation}
P_{CMAX} = P_{MAX} - P_{MIN} .
\end{equation}

 
\subsection{MUSE: Definitions}
\noindent
\textbf{Transmitter-occupancy:}
\noindent
We define transmitter-occupancy as the amount of spectrum consumed by a transmitter \textit{at a point} in terms of RF-power occupied at the point.

\noindent
\textbf{Receiver-liability:}
\noindent
We define receiver-liability as the amount of spectrum consumed by a receiver \textit{at a point} in terms of the constraint imposed on the RF-power that can be exercised at the point by a potential or an existing transmitter. Thus, it represents \textit{liability to the receiver} in order to protect the receiver from harmful interference.

\noindent
\textbf{Discretized Spectrum-space:}
\noindent
The spectrum consumed by a transmitter or a receiver is \textit{continuous} in the space, time, and frequency dimensions. In order to facilitate characterization and quantification of the use of the spectrum within a geographical region, we \textit{divide} the total spectrum-space into discrete units and characterize the spectrum consumed by the transmitters and receivers in the unit spectrum-spaces. We refer to this discretized view of the spectrum in the space, time, and frequency dimensions as \textit{discretized spectrum-space}.

\noindent
\textbf{A unit spectrum-space:}
\noindent
A unit spectrum-space represents the spectrum within an unit area, in a unit time-quanta, and a unit frequency band. 

\noindent
\textbf{RF-entity:}
We use an RF-entity as a generic term for an entity exercising spectrum-access. A RF-entity may represent an individual transmitter, an individual receiver, a RF-Link, a RF-network, or a RF-system.

\begin{figure*}[htbp!]
\centering
{\includegraphics [width=0.94\textwidth, angle=0] {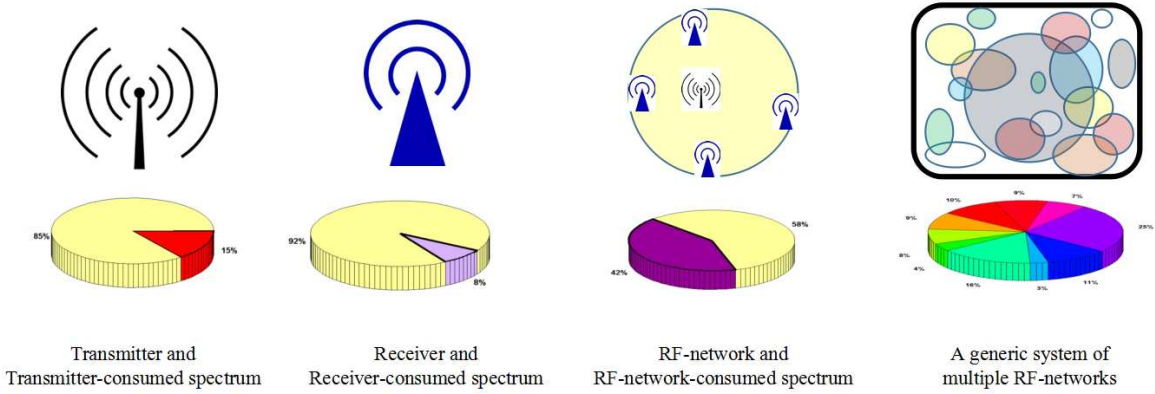}}
\setlength{\abovecaptionskip}{-9pt}
\caption{RF-entities and associated spectrum consumption spaces. The figure shows various entities within a generic system of wireless services sharing RF-spectrum: a transmitter, a receiver, and a RF-network. The rightmost picture shows a generalized  spectrum sharing scenario with multiple \textit{spatially-overlapping} heterogeneous wireless services sharing spectrum in the time, space, and frequency dimensions. The generalized topology emphasizes on the significance of spectrum sharing among heterogeneous wireless services without defining spatial, temporal, and spectral boundaries.} 
\label{fig:nw_term_illn}
\end{figure*}

\noindent
\textbf{A spectrum consumption space:}
\noindent
A spectrum consumption space captures the spectrum consumption by a RF-entity in the discretized spectrum-space. The unit of a spectrum consumption space is $Wm^2$. Figure~\ref{fig:nw_term_illn} shows different RF-entities and the associated spectrum consumption spaces.

\noindent
\textbf{The total spectrum-space:}
\noindent
The total spectrum-space represents the spectrum in the space, time, and frequency dimensions within a geographical region of interest. Let the geographical region be discretized into $\hat{A}$ unit-regions, $\hat{B}$ unit-frequency-bands, and $\hat{T}$ unit-time-quanta.  Thus, the total spectrum-space is given by
\begin{equation}
\label{eq:p1totalspectrum}
\Psi_{Total} = P_{CMAX}\ \hat{T} \hat{A} \hat{B} .
\end{equation}

We identify following attributes with a unit-spectrum-space to characterize and quantify spectrum consumption spaces. 

\noindent
\textbf{Unit-spectrum-space Occupancy:}
\noindent
We define unit-spectrum-space occupancy as the amount of spectrum consumed by all the transmitters in a unit-spectrum-space.

\noindent
\textbf{Unit-spectrum-space Liability:}
\noindent
We define unit-spectrum-space liability as the amount of spectrum consumed by all the receivers in a unit-spectrum-space.

\noindent
\textbf{Unit-spectrum-space Opportunity:}
\noindent
We define unit-spectrum-space opportunity as the amount of spectrum available for consumption in a unit-spectrum-space.

\subsection{Quantifying Spectrum Consumption} 
\noindent
\textbf{Transmitter-occupancy} \\
\noindent
The power received from a transmitter $t_n$ at a point $\rho$ in the spatial dimension is given by 
\begin{equation}
\label{eq:txocpt}
P_{r_{\rho}}(t_n) = P_{t_n} min\Big\{1, L({d(t_n,\rho)}^{-\alpha})\Big\} ,
\end{equation}
where $P_{t_n}$ is the transmit power of the transmitter and $d(t_n,\rho)$ is the distance between the transmitter $t_n$ and the point $\rho$ in the space. ${\alpha}$ is the path-loss exponent and $L({d(t_n,\rho)}^{-\alpha})$ denotes the path-loss factor. Thus, (\ref{eq:txocpt}) represents transmitter-occupancy of $t_n$ at the point $\rho$ in the geographical region.

\noindent
\textbf{Spectrum-occupancy} \\
\noindent
The aggregate power received at a point $\rho$ is given by
\begin{equation}
\label{eq:spocpt}
\bar{P}_{\rho} = \sum_{n}	P_{r_\rho}(t_n) + W_{\rho} ,
\end{equation}
where $W_{\rho}$ is the average ambient noise power at $\rho$. Thus, (\ref{eq:spocpt}) represents the spectrum-occupancy at the point $\rho$ in the geographical region. 

\noindent
\textbf{Unit-spectrum-space occupancy} \\
\noindent
Let us consider a unit-spectrum-space defined by unit-region $\chi$, time-quanta ${\tau}$, frequency-band $\nu$. We define unit-spectrum-space occupancy, $\omega(\chi, \tau, \nu)$, as the spectrum occupancy at the \textit{sample point} $\rho_0 \in \chi$, in the frequency band $\nu$, at an instant within the time-quanta ${\tau}$. Therefore, 
\begin{equation}
\label{eq:ussoc}
\omega(\chi, \tau, \nu) = \bar{P}_{\rho_0} .
\end{equation}
The unit for unit-spectrum-space occupancy is $W$.

\noindent
\textbf{Receiver interference-margin} \\
Let $r_{n,m}$ be the $m^{th}$ receiver of the $n^{th}$ RF-link. The amount of interference power receiver $r_{n,m}$ can tolerate, that is the interference-margin for $r_{n,m}$, is given by 
\begin{equation}
\breve{P}_{r_{n,m}} = \frac{P_{r_{n,m}}(t_n)}{\beta_{n,m}} - W_{r_{n,m}} .
\end{equation}
The unit of interference-margin is $W$.

We can view interference-margin $\breve{P}_{r_{n,m}}$ as the upper-bound on the transmit power of an interferer at a \textit{spatial separation of zero}. We characterize the limit on the interference power at a point $\rho$ in space in terms of the receiver-imposed upper bound on the interference power.
\begin{equation}
\label{eq:riub}
\acute{I}(r_{n,m}, \rho) = \breve{P}_{r_{n,m}} min\{1, L(d(\rho, r_{n,m})^{\alpha})\} ,
\end{equation}
where $d(\rho, r_{n,m})$ is the distance between the receiver $r_{n,m}$ and the point $\rho$ in the space. We note that the receiver imposed constraint on the interference power increases with increasing separation.

\noindent
\textbf{Receiver-imposed interference-opportunity} \\
Let $\breve{I}(r_{n,m}, \rho)$ denote the \textit{proportional aggregate interference power}\footnote{that is, the aggregate RF-power received at $\rho$ from all the interference sources for the receiver $r_{n,m}$.} seen at a distant point $\rho$; then the \textit{interference opportunity} imposed by this receiver at $\rho$ is given by the difference between the upper bound on the interference power and the proportional aggregate interference power.
\begin{equation}
\label{eq:spoppt}
\ddot{I}(r_{n,m}, \rho) = \acute{I}(r_{n,m}, \rho) -  \breve{I}(r_{n,m}, \rho) .
\end{equation}
We note that when $\ddot{I}(r_{n,m}, \rho)$ is negative, the receiver $r_{n,m}$ is experiencing harmful interference. 

\noindent
\textbf{Spectrum-opportunity} \\
By combining the limits on the maximum interference power imposed by all the receivers, from all the RF-links in the system, we obtain net interference-opportunity at a point $\rho$ as
\begin{equation}
\label{eq:niub}
\bar{I}_{\rho}	= \min_n (\min_m (\ddot{I}(r_{n,m}, \rho))) .
\end{equation}
We term the net interference-opportunity at a point as \textit{spectrum-opportunity}. 

\noindent
\textbf{Unit-spectrum-space opportunity} \\
We define unit-spectrum-space opportunity, $\gamma(\chi, \tau, \nu)$, as the spectrum-opportunity at the sample point $\rho_0 \in \chi$, in frequency band $\nu$, at an instant within the time-quanta ${\tau}$. Therefore, 
\begin{equation}
\label{eq:ussop}
\gamma(\chi, \tau, \nu)	= \bar{I}_{\rho_0} .
\end{equation}
The unit for unit-spectrum-space opportunity is $W$. 

\noindent
\textbf{Unit-spectrum-space liability} \\
We obtain unit-spectrum-space liability, that is the spectrum consumed by all the receivers in a unit-spectrum-space by subtracting the unit-spectrum-space occupancy and unit-spectrum-space opportunity from the maximum spectrum-consumption. Therefore, 
\begin{equation}
\label{eq:ussrl}
\phi(\chi, \tau, \nu) = P_{CMAX} - (\omega(\chi, \tau, \nu) + \gamma(\chi, \tau, \nu)) .
\end{equation}
The unit for unit-spectrum-space liability is $W$. 

Characterizing the spectrum consumed by a RF-entity at a point enables characterizing the spectrum consumption space associated with the RF-entity. In this regard, we characterize the spectrum consumed by an individual transceiver in a unit-spectrum-space.

\noindent
\textbf{Receiver-liability} \\
We get receiver-liability, that is the amount of spectrum consumed by an \textit{individual} receiver at a point by subtracting the aggregate transmitter-occupancy and the interference-opportunity caused by the receiver from the maximum spectrum-consumption. Therefore, 
\begin{equation}
\label{eq:rxrlpt}
L_{\rho}(r_{n,m}) = P_{CMAX} - (\bar{P}_{\rho} + \ddot{I}(r_{n,m}, \rho)) .
\end{equation}
The unit of receiver-liability is $W$.

\noindent
\textbf{Transmitter-occupancy in a unit-spectrum-space} is given by transmitter-occupancy at the sample point $\rho_0 \in \chi$, in frequency band $\nu$, at an instant within the time-quanta ${\tau}$. Therefore, 
\begin{equation}
\label{eq:txspoc}
\omega_{t_n}(\chi, \tau, \nu) = P_{r_{\rho_0}}(t_n) .
\end{equation}

\noindent
\textbf{Receiver-liability in a unit-spectrum-space} is given by receiver-liability at the sample point $\rho_0 \in \chi$, in frequency band $\nu$, at an instant within the time-quanta ${\tau}$. Therefore, 
\begin{equation}
\label{eq:rxsprl}
\phi_{r_{n,m}}(\chi, \tau, \nu) = L_{\rho_0}(r_{n,m})  .
\end{equation}
%
\subsection{Quantifying a Spectrum Consumption Space}
A spectrum consumption space associated with a RF-entity is quantified by aggregating the spectrum consumed by the RF-entity across all the unit-spectrum-spaces within a geographical region. We identify a few spectrum consumption spaces towards maximizing the use of spectrum in Table I and quantify these spectrum spaces in this subsection.

\vspace{\baselineskip}
\begin{table}[h!t!b!p!]
\setlength{\abovecaptionskip}{-8pt}
\caption{Example spectrum consumption spaces} 
\centering
\begin{tabular}{|p{3cm}|p{5cm}|p{7.5cm}|}
\hline
Spectrum Consumption Space & Description & Significance\\
\hline
 Transmitter-consumed spectrum & It represents the spectrum consumed by a specified transmitter. & It can be used in the context of \textit{defining and enforcing} spectrum-access rights for a single transmitter\\
 Receiver-consumed spectrum & It represents the spectrum consumed by a specified receiver.& It can be used in the context of \textit{defining and enforcing} spectrum-access rights for a single receiver\\
 Utilized-spectrum  & It represents the spectrum consumed by all the transmitters in the system.&It can be used in the context of analysis and optimization of the spectrum consumed by transmitters.\\
 Forbidden-spectrum  & It represents the spectrum consumed by all the receivers in the system.&It can be used in the context of analysis and optimization of the spectrum consumed by receivers.\\
 Available-spectrum  & It represents the spectrum not consumed by all the transmitters and receivers in the system and therefore \textit{available}\tablefootnote{In a spectrum sharing scenario, the spectrum-sharing policy defines certain constraints which determine what spectrum can be exercised for shared-access. We distinguish the spectrum implied available by a spectrum-sharing policy, that is, the implied-available spectrum, from the available-spectrum.} for consumption.& It can be used in the context of analysis of the potential of spectrum sharing and for assigning spectrum-access footprints.\\
\hline	
\end{tabular}
\end{table}

\noindent
\textbf{Transmitter-consumed spectrum:}
The spectrum consumed by a transmitter within a geographical region is obtained by aggregating transmitter-occupancy across the unit-spectrum-spaces. Therefore, 
\begin{equation}
\label{eq:txsput}
\Omega(t_n) = \sum_{k=1}^{\hat{B}} \sum_{j=1}^{\hat{T}} \sum_{i=1}^{\hat{A}} {\omega}_{t_n}(\chi_i, \tau_j, \nu_k) .
\end{equation}

\noindent
\textbf{Receiver-consumed spectrum:}
The spectrum consumed by a receiver within a geographical region is obtained by aggregating receiver-liability across the unit-spectrum-spaces. Therefore, 
\begin{equation}
\label{eq:rxsput}
\Phi(r_{n,m}) = \sum_{k=1}^{\hat{B}} \sum_{j=1}^{\hat{T}} \sum_{i=1}^{\hat{A}} {\phi}_{r_{n,m}}(\chi_i, \tau_j, \nu_k) .
\end{equation}

\noindent
\textbf{Utilized-spectrum ($\Psi_{utilized}$):}
We define \textit{utilized-spectrum} as the spectrum consumed by all the transmitters within a geographical region. Utilized-spectrum is obtained by summing the unit-spectrum-space occupancy across all the unit-spectrum-spaces. Therefore, 
\begin{equation}
\label{eq:agsput}
\Psi_{utilized} = \sum_{k=1}^{\hat{B}} \sum_{j=1}^{\hat{T}} \sum_{i=1}^{\hat{A}} {\omega}(\chi_i, \tau_j, \nu_k) .
\end{equation}

\noindent
\textbf{Forbidden-spectrum ($\Psi_{forbidden}$):}
We define \textit{forbidden-spectrum} as the spectrum consumed by all the receivers within a geographical region. The forbidden-spectrum is quantified by aggregating unit-spectrum-space reliability across all the unit-spectrum-spaces. Therefore, 
\begin{equation}
\label{eq:agspfb}
\Psi_{forbidden} = \sum_{k=1}^{\hat{B}} \sum_{j=1}^{\hat{T}} \sum_{i=1}^{\hat{A}} {\phi}(\chi_i, \tau_j, \nu_k) .
\end{equation}

\noindent
\textbf{Available-spectrum ($\Psi_{available}$):}
We define available-spectrum as the spectrum not consumed transmitters and receivers and therefore \textit{available} for consumption. The available-spectrum within a geographical region is obtained by summing unit-spectrum-space opportunity across all the unit-spectrum-spaces. Therefore, 
\begin{equation}
\label{eq:agspav}
\Psi_{available} = \sum_{k=1}^{\hat{B}} \sum_{j=1}^{\hat{T}} \sum_{i=1}^{\hat{A}} {\gamma}(\chi_i, \tau_j, \nu_k) .
\end{equation}

For \textit{completeness}\footnote{In fact, this relationship follows from the definition of unit-spectrum-space reliability.}, we express the relationship between these spectrum consumption spaces.
The spectrum consumption in a unit-spectrum-space is specified in terms of the unit-spectrum-space occupancy, unit-spectrum-space opportunity, and unit-spectrum-space liability. From (\ref{eq:ussrl}),
\begin{equation}
{\omega}(\chi, \tau, \nu) + {\phi}(\chi, \tau, \nu)  + {\gamma}(\chi, \tau, \nu)	= P_{CMAX} .
\end{equation}
Summing over all the $\hat{A}$ unit-regions in the geographical-region, $\hat{B}$ frequency-bands, $\hat{T}$ unit-time quanta, we get following relation between utilized-spectrum, forbidden-spectrum, and available-spectrum.
\begin{equation}
\label{eq:p1e_sp_csm_reln}
\Psi_{utilized} + \Psi_{forbidden} + \Psi_{available}  = \Psi_{Total} .
\end{equation}

\noindent
\textbf{Quantifying other spectrum consumption spaces}\\
One can identify a spectrum consumption space with regards to the desired objective and quantify the spectrum consumption space to facilitate analysis and optimization. For example, one can quantify the harmful interference caused by a single transmitter to the cochannel receivers. This can be useful in terms of regulation of a spectrum-access policy.

\subsection{Characterizing and Quantifying Performance of the Spectrum Management Functions}

Similar to characterization of the use of spectrum in terms of spectrum consumption spaces, we characterize the performance of a spectrum management function in terms of the \textit{spectrum management function (SMF) spaces}.

Let us consider an attribute, $\theta$, characterizing the performance of a spectrum management function at a point. For example, in case of spectrum recovery, the error in the estimated unit-spectrum-space opportunity may represent the performance of spectrum recovery or in case of spectrum exploitation, the amount of spectrum not exploited in a unit-spectrum-space can capture weaknesses of a spectrum exploitation mechanism. Table III describes the SMF spaces associated with spectrum sharing, spectrum recovery, and spectrum exploitation functions.

We characterize the SMF attribute $\theta$ in a \textit{unit-spectrum-space} defined by unit-region $\chi$, time-quanta ${\tau}$, frequency-band $\nu$ in terms of the SMF attribute at the sample point $\rho_0 \in \chi$, in the frequency band $\nu$, at an instant within the time-quanta ${\tau}$. Therefore,
\begin{equation}
\label{eq:usssmf}
\theta(\chi, \tau, \nu) = \theta_{\rho_0} .
\end{equation}

The SMF space within a geographical region is obtained by summing the SMF attribute across the unit-spectrum-spaces. Therefore, 
\begin{equation}
\label{eq:smfag}
\Theta = \sum_{k=1}^{\hat{B}} \sum_{j=1}^{\hat{T}} \sum_{i=1}^{\hat{A}} {\theta}(\chi_i, \tau_j, \nu_k) .
\end{equation}
%
\section{MUSE: Illustration and Discussion}
We start illustration of the methodology with an abstract view of the use of spectrum at a single point. The total spectrum at a point is determined by $P_{MAX}$ and $P_{MIN}$. 
\begin{itemize}
	\item If there are no transmitters and receivers in the system, transmitter-occupancy and receiver-liability at this point are zero; the spectrum-opportunity will be maximum, that is, $P_{CMAX}$. The spectrum-opportunity represents maximum RF-power that can be used by a future transmitters while ensuring non-harmful interference at the receivers. This scenario is captured by the leftmost bar.
	\item If we add a pair of transmitter and its receiver, we can observe nonzero transmitter-occupancy and receiver-liability. A key thing to note is receiver-liability being the limit on the maximum RF-power at a point, grows from $P_{MAX}$ towards $P_{MIN}$. Thus, higher the minimum SINR for successful reception, higher is the receiver-liability. The transmitter-occupancy and receiver-liability shape the spectrum-opportunity at a point. The middle bar shows this scenario and we can observe that the spectrum-opportunity has reduced due to the constraint imposed by receiver.
	\item As more and more transceivers are added in the system, the spectrum-opportunity goes on reducing. This scenario is shown in the rightmost bar. In this case, the spectrum-occupancy captures the aggregate value of the transmitter occupancy from the individual transmitters. With regards to receivers, different receivers impose a different constraint on the RF-power sourced from the point. The effective constraint at this point is determined by the receiver having the highest receiver-liability at the point. 
\end{itemize}
\begin{figure}[htbp!]
\centering
{\includegraphics [width=0.6\textwidth, angle=0] {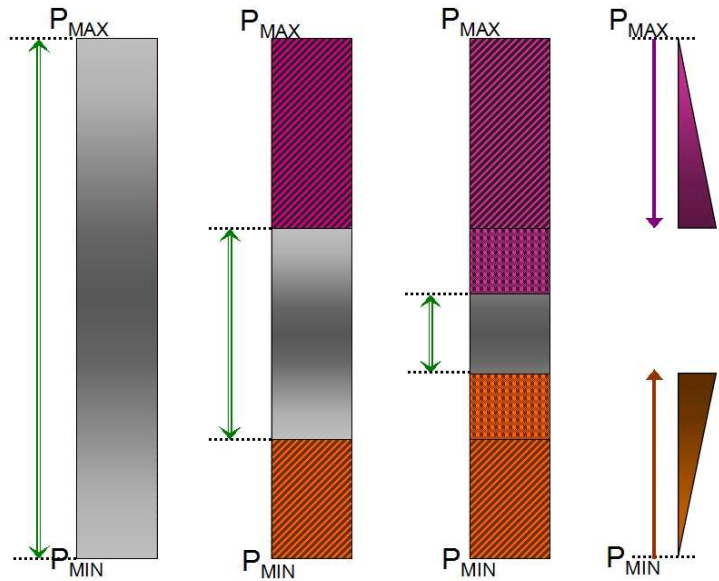}}
\setlength{\abovecaptionskip}{-5pt}
\caption{The use of spectrum at a point. The leftmost bar captures the maximum ($P_{MAX} - P_{MIN}$) spectrum-opportunity (shown with green double arrow) at a point when no transceivers are present. The middle bar shows the spectrum consumed by a transmitter and its receiver. The rightmost bar shows the spectrum consumed by two pairs of transceivers. Here, we note that the spectrum-occupancy grows from $P_{MIN}$ towards $P_{MAX}$ while spectrum-liability representing a constraint on the occupiable RF-power grows from  $P_{MAX}$ towards  $P_{MIN}$. The spectrum-opportunity goes on reducing as the transceivers consume more and more of the spectrum at a point.}
\label{fig:qus}
\end{figure}

Next, we consider a simplistic setup and illustrate the spectrum consumed by the transceivers at an arbitrary point in the system. Let us consider a 4.3 km $\times$ 3.7 km geographical region. We assume, $P_{MAX}$ is 1 W or 30 dBm; $P_{MIN}$ is $-200$ dBm. Ambient noise floor is assumed to be -106 dBm (for channel bandwidth of 6 MHz). Distance dependent path-loss model is used with path-loss exponent of 3.5. The minimum desired SINR at receiver for successful reception, $\beta$, is assumed to be 3 dB. A transmitter is positioned at (1000, 1200), receiver is positioned at (1000, 2100), and spectrum consumption is quantified at an arbitrary point, (2250, 1800). The scenario I in Table II shows the spectrum consumption at this point in terms of transmitter-occupancy, receiver-liability, and spectrum-opportunity.
\begin{table}[h!b!p!]
\setlength{\abovecaptionskip}{-5pt}
\caption{Spectrum consumption at a point under three scenarios.} 
\centering
\begin{tabular}{|p{0.5cm}|p{1.5cm}|p{2cm}|p{2cm}|p{2cm}|p{2cm}|p{3cm}|}
\hline
S/N & Transmit Power & Receiver position & Receiver SINR & Transmitter-Occupancy & Spectrum-Opportunity & Receiver-Liability\\
\hline
 I & -24 dBm & (1000, 2100) & 12.0 dB & -132.6 dBm & 11.84 dBm & 29.93 dBm (984 mW)\\
 II & 6 dBm & (1000, 2100) & 42.0 dB & -102.6 dBm & 30 dBm & -200 dBm (0 mW)\\
 III & 6 dBm & (1000, 2500) & 17.5 dB & -102.6 dBm & 19.0 dBm& 29.64 dBm (920 mW)\\
\hline
\end{tabular}
\end{table}

In scenario II, we change the transmitter power from -24 dBm  to 6 dBm; accordingly, the spectrum consumption by the transmitter at point (2250, 1800) changes from -132.6 dBm to -102.6 dBm. With regards to the spectrum consumption by the receiver, we observe that SINR at the receiver is significantly improved and consequently the tolerance for interference is improved. Thus, the receiver can withstand interference of 30 dBm generated from position (2250, 1800) without getting harmfully interfered. Since, the spectrum-opportunity is maximum (30 dBm or $P_{MAX}$) in case II, the spectrum consumption by the receiver is minimal ($-200$ dBm or $P_{MIN}$). 

In Scenario III, we move the receiver farther from its transmitter; this results in reduced SINR and consequently reduced tolerance to interference. Thus, the spectrum-opportunity caused by the receiver at (2250, 1800) is lowered from 30 dBm to 19 dBm and the spectrum consumed by the receiver increases to 920 mW. 
\subsection{Quantification of a Spectrum Consumption Space}
After characterizing spectrum consumption at an arbitrary point, we move to characterizing a spectrum consumption space within a geographical region of interest. Let us consider a 4.3 km $\times$ 3.7 km geographical region with 676 hexagonal unit regions with each side 100 m long. Let the maximum RF-power at a point, $P_{MAX}$ in the unit regions be 30 dBm , that is 1 W. Let $P_{MIN}$ be $-200$ dBm. Let us consider 6 MHz spectral range as unit bandwidth and a 10 second time period as unit time. In this scenario, the maximum spectrum consumption in the geographical region, in a 6 MHz spectral band, in a 10 second time period is 676 $Wm^2$ as given by (\ref{eq:p1totalspectrum}). We model the propagation conditions by distance dependent path-loss model with the path-loss exponent is 3.5. 

\subsubsection{Spectrum Consumed by a Transmitter}
\begin{figure}[htbp]
\centering
{\includegraphics [width=0.48\textwidth, angle=0] {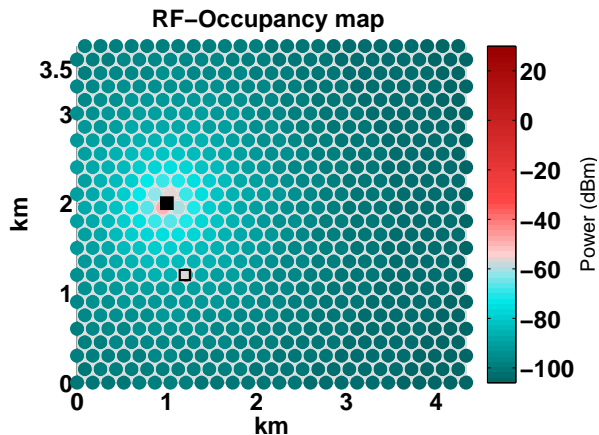}}
\setlength{\abovecaptionskip}{-11pt}
\caption{Spectrum consumption space of an individual transmitter. The figure shows spatial distribution of the transmitter-occupancy in the unit-spectrum-spaces within a geographical region. Thus, it captures the spectrum consumed by a transmitter within the geographical region. The transmitter is shown by a solid square and the receiver is shown by a non-solid square.} 
\label{fig:L201}
\end{figure}

First, we will look into the spectrum consumption space for an individual transmitter. Figure~\ref{fig:L201} illustrates the spectrum consumed by a transmitter within a geographical region according to (\ref{eq:txsput}). The transmitter is located at (1000, 2000) and is exercising omnidirectional transmission with transmit power of 15 dBm. The spectrum consumed by the transmitter is 1.8 x $10^{-8}$ $Wm^2$ (2.7 x $10^{-9}$ \% of the total spectrum space). 
%
\subsubsection{Spectrum Consumed by a Receiver}
\begin{figure}[htbp]
\centering
{\includegraphics [width=0.48\textwidth, angle=0] {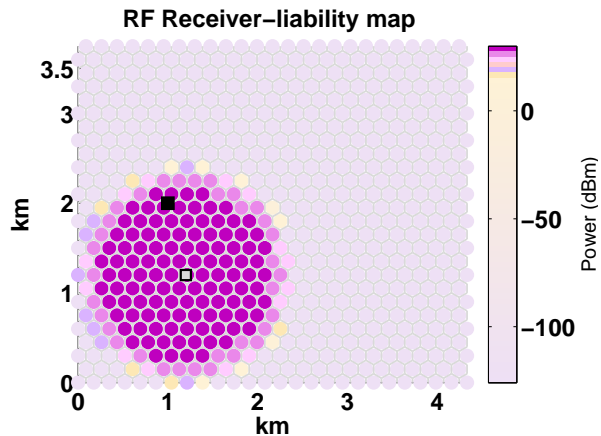}}
\setlength{\abovecaptionskip}{-11pt}
\caption{Spectrum consumption space of an individual receiver. The figure shows the spatial distribution of receiver-liability in the unit-spectrum-spaces within a geographical region. Thus, it captures the spectrum consumed by a receiver within the geographical region. The transmitter is shown by a solid square and the receiver is shown by a non-solid square.} 
\label{fig:L202_RL}
\end{figure}

Figure~\ref{fig:L202_RL} illustrates the spectrum consumption space for a receiver according to (\ref{eq:rxsput}). The receiver is located at (1200, 1200) and is exercising omnidirectional reception requiring minimum SINR of 6 dB and the actual experienced SINR at the receiver is 33 dB. We note that as the distance from a receiver increases, a cochannel transmitter can exercise higher transmission power. Thus, the liability for ensuring non-harmful interference to the receiver goes down with the distance from the receiver. The spectrum consumed by the receiver is 112.4 $Wm^2$ (16.6\% of the total spectrum space).
%
\subsubsection{Available-spectrum} 
\begin{figure}[htbp]
\centering
{\includegraphics [width=0.52\textwidth, angle=0] {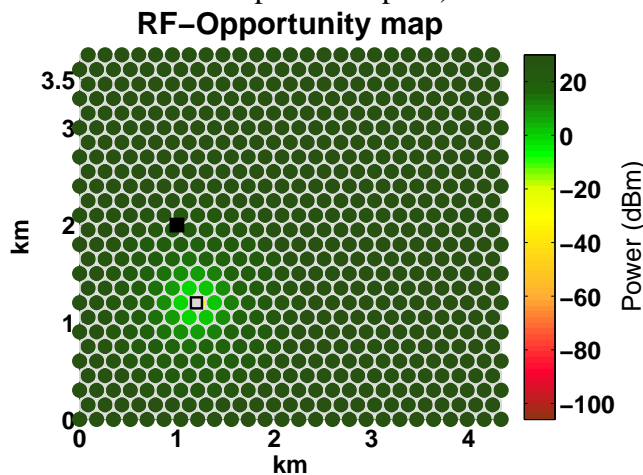}}
\setlength{\abovecaptionskip}{-11pt}
\caption{Available-spectrum within a geographical region. The figure shows the spatial distribution of unit-spectrum-space opportunity within a geographical region. Thus, it captures the available-spectrum within the geographical region. The transmitter is shown by a solid shape and the receiver is shown by a non-solid shape. We observe that the unit-spectrum-space opportunity near the receiver is lower in order to ensure non-harmful interference at the receiver.} 
\label{fig:L202_OP}
\end{figure}

The spectrum space not consumed by the transmitters and receivers is the available-spectrum within a geographical region. Figure~\ref{fig:L202_OP} depicts spatial distribution of unit-spectrum-space opportunity given by (\ref{eq:agspav}) for the above topology. The available-spectrum space is 563.6 $Wm^2$ (83.4\% of the total spectrum space).

The spectrum consumed by a RF-entity is the sum of the spectrum consumed by all the transmitters and receivers within the RF-entity. Thus, the spectrum consumed by the RF-link in this case is 112.4 $Wm^2$. The spectrum consumed by a RF-link has been considered as a parameter for scheduling RF-links in order to minimize spectrum consumption and improve the performance of scheduling in \cite{oms3_cf1}.
%
\subsection{Quantification of a Spectrum Management Space}
The performance of spectrum sharing depends primarily on the spectrum-sharing policy which defines what spectrum can be \textit{considered available} for exploitation, the \textit{optional}\footnote{The management of spectrum may vary across different spectrum sharing models. Market based approach to spectrum sharing presumes a spectrum pool while overlay approach requires recovering the underutilized spectrum.} spectrum recovery function which defines how efficiently the available spectrum is recovered, and the spectrum exploitation function which influences how efficiently the recovered available spectrum is assigned for satisfying spectrum-access requests. In this regard, Table III identifies a few example spectrum management spaces. 
\begin{table}[h!b!p!]
\setlength{\abovecaptionskip}{-5pt}
\caption{Example spectrum management spaces} 
\centering
\begin{tabular}{|p{2.5cm}|l|p{8cm}|}
\hline
Functionality & Spectrum Management Space & Description\\
\hline
 Spectrum Sharing & Implied-Available Spectrum Space & It represents the portion of available spectrum implied accessible under the constraints imposed by a spectrum-sharing policy.\\
   & Implied-Guard Spectrum Space & It represents the portion of available spectrum that has been (usually intentionally) treated as not available spectrum.\\
   & Implied-Incursed Spectrum Space  & It represents the portion of non available spectrum that has been erroneously treated as available spectrum.\\
\hline	
 Spectrum Recovery & Recovered-Available Spectrum Space & It represents the portion of implied-available spectrum that has been recovered for exploitation.\\
   & Lost-Available Spectrum Space & It represents the portion of implied-available spectrum that has been erroneously treated as not available for exploitation.\\
   & Potentially-incursed Spectrum Space & It represents the portion of non available spectrum that has been erroneously treated as available for exploitation.\\
\hline	
 Spectrum Exploitation & Exploited-Available Spectrum Space & It represents the portion of the recovered-available spectrum consumed by transmitters and receivers.\\
   & Unexploited-Available Spectrum Space & It represents the portion of the recovered-available spectrum not consumed by transmitters and receivers.\\
   & Incursed Spectrum Space & It represents the portion of the non available spectrum consumed by transmitters and receivers.\\
\hline
\end{tabular}
\end{table}

\begin{figure}[htbp!]
\centering
{\includegraphics [width=0.56\textwidth, angle=0] {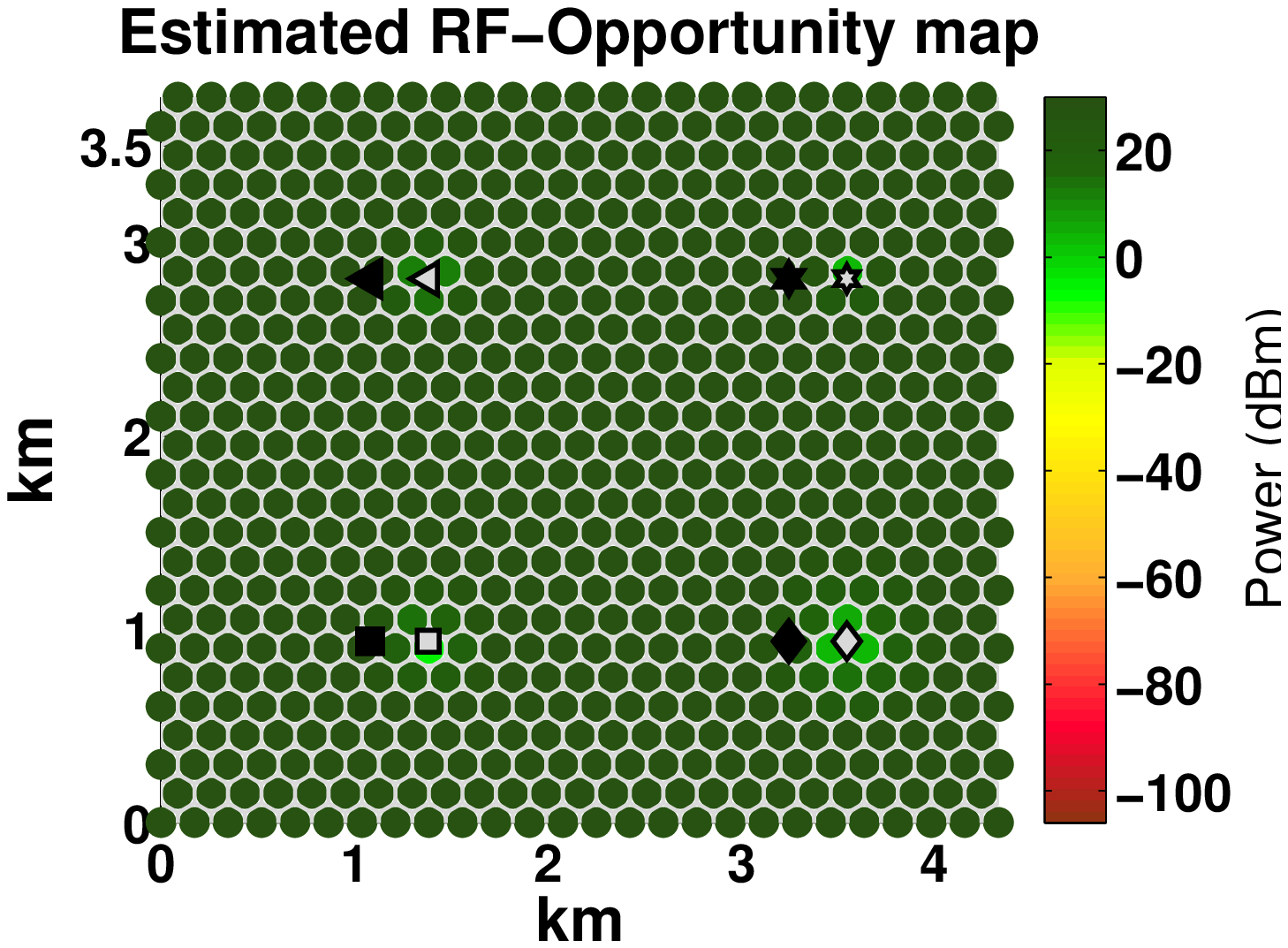}}
{\includegraphics [width=0.48\textwidth, angle=0] {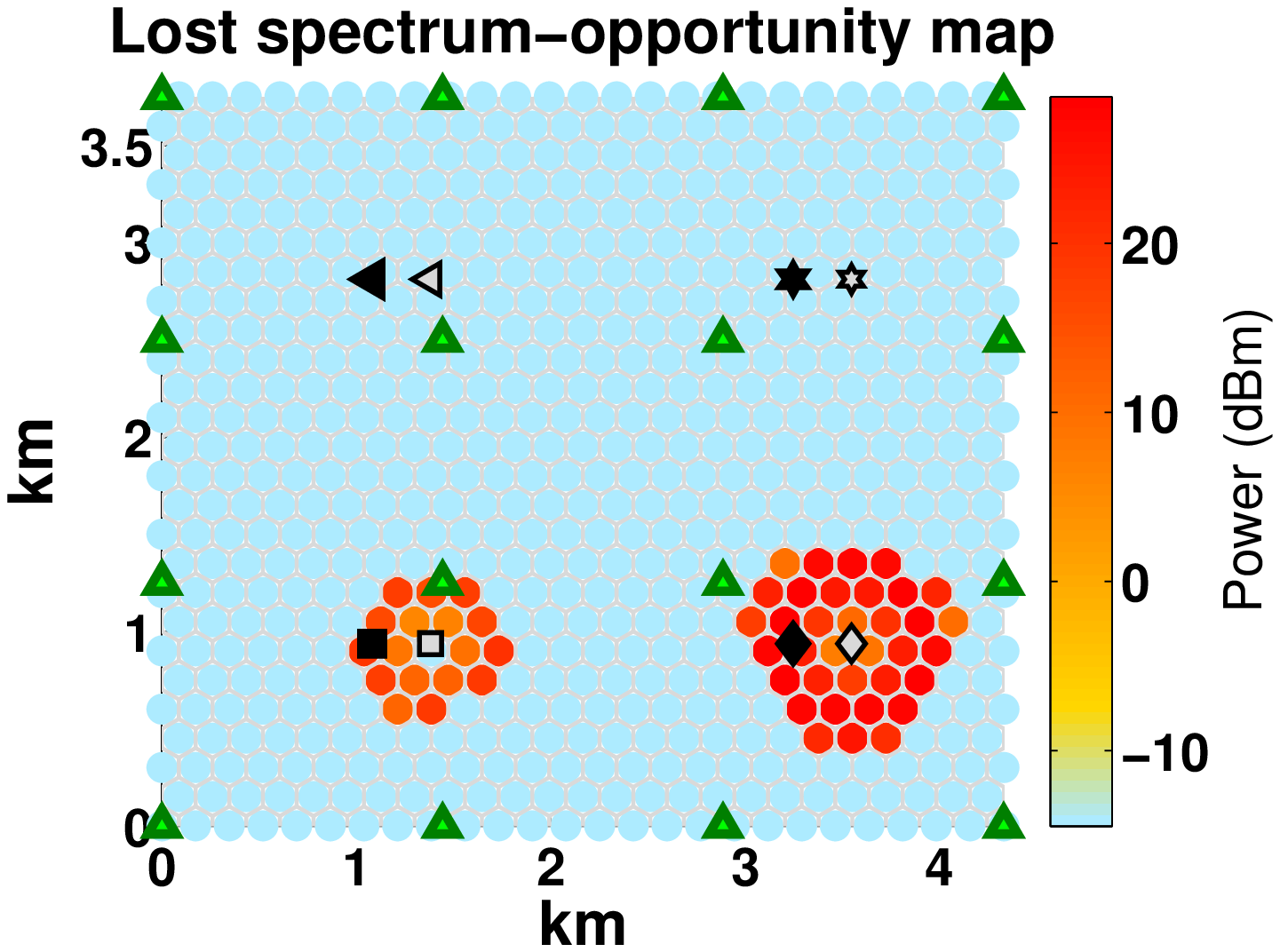}}
{\includegraphics [width=0.48\textwidth, angle=0] {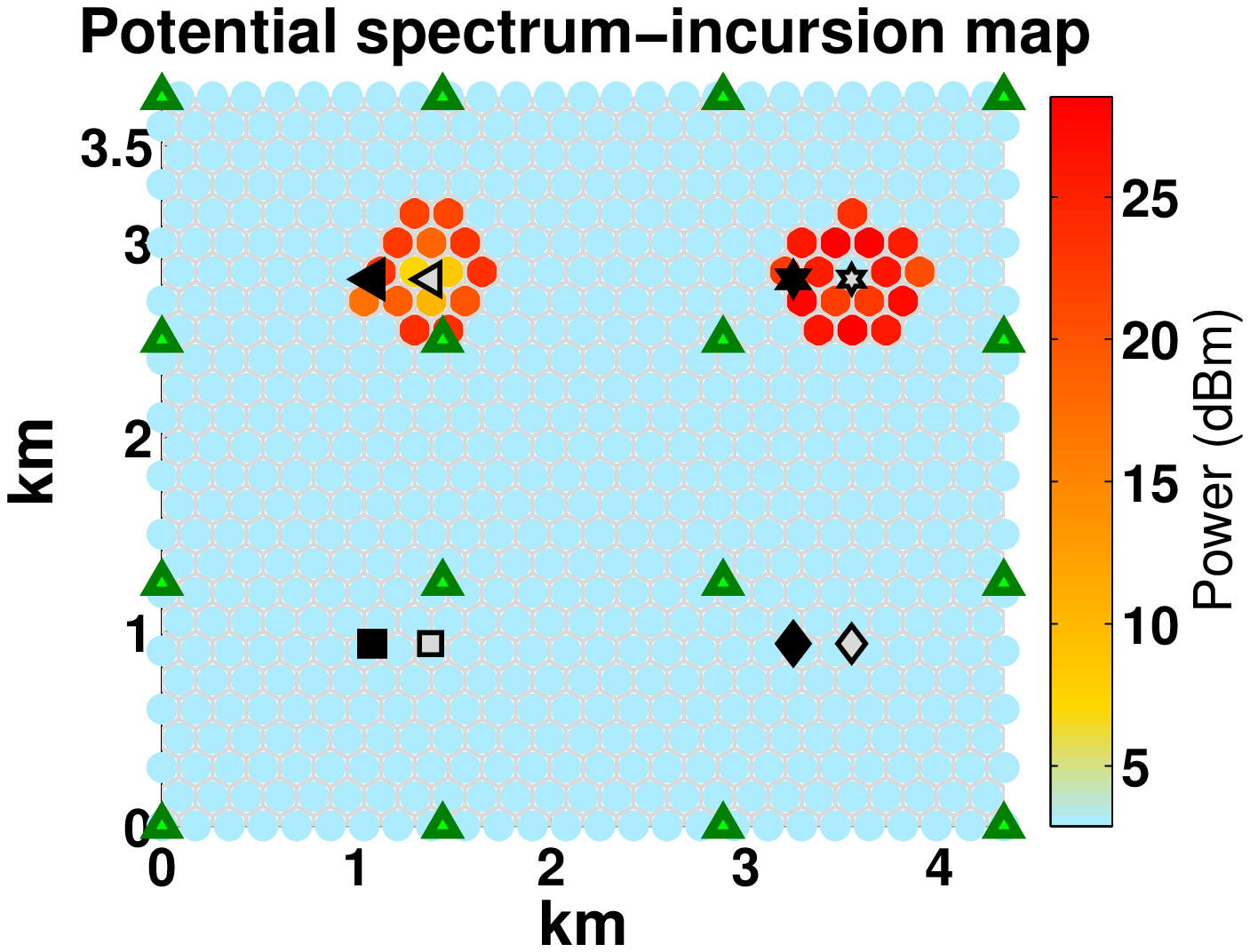}} 
\setlength{\abovecaptionskip}{-11pt}
\caption{Performance analysis of recovering the available-spectrum. A dedicated RF-sensor network with 16 RF-sensors estimates the unit-spectrum-space opportunities exploiting signal-cyclostationarity across the unit-spectrum-spaces within a geographical region \cite{oms4_sce}. The errors in the detection, geolocation, and transmit-power estimation result into lost-available spectrum and potentially-incursed spectrum. The top plot characterizes the estimated unit-spectrum-space opportunities. The bottom two plots capture the spatial distribution of lost-spectrum-opportunity and potential-spectrum-incursion within the geographical region. The RF-sensors are shown by upward-pointing triangle markers. The other 4 markers represent 4 transceiver pairs; a solid marker shows a transmitter and an unfilled marker shows a receiver.} 
\label{fig:sphrvperf}
\end{figure}

Quantifying these spectrum management spaces enables comparison, analysis, and optimization of spectrum sharing. Let us consider the recovery of the underutilized spectrum by estimating the unit-spectrum-space opportunities within a geographical region using a dedicated RF-sensor network. The RF-sensors sense the RF-environment in order to detect the presence of cochannel transmitters, geolocate the transmitters, and estimate the transmit-power of the transmitters. A missed-detection, false-positive, an error in geo-location implies error in the estimated spectrum-opportunity. A negative error implies spectrum opportunity is lost in the unit-spectrum-space and a positive error may potentially lead to harmful interference. 
Figure~\ref{fig:sphrvperf} shows the performance of spectrum recovery in term of estimation of unit-spectrum-space opportunity given by (\ref{eq:ussop}). We observe that 630.7 W (99.3 \% of the total spectrum) of the available-spectrum within the geographical region has been recovered; 12.7 W (2 \% of the total spectrum) of the available-spectrum has been lost and  8.4 W (1.3 \% of the total spectrum) of the not available-spectrum has been erroneously considered available for exploitation \cite{oms4_sce}.  

\subsection{Considerations while Applying MUSE}
%
\subsubsection{Unit-spectrum-space Dimensions}
The granularity of spectrum sharing identifies the smallest portion of spectrum-space for which spectrum-access rights could be defined and enforced.  In this regard, a unit spectrum-space represents the lowest granularity of spectrum sharing. Thus, the granularity of spectrum sharing plays a key role in determining the sampling rate in the space, time, and frequency dimensions.
In favor of standardization, an alternate perspective could be choosing the unit-spectrum-space dimensions; thus, the unit-spectrum-space granularity could determine the spectrum-sharing granularity. In this case, we suppose the \textit{population-density} and the \textit{propagation environment characteristics} can play a key role in determining the spatial granularity of a unit-spectrum-space. The temporal granularity for a unit-spectrum-space can be considered to depend upon the traffic characteristics. The transceiver technology and its frequency agility would typically drive the granularity of a unit-spectrum-space in the frequency dimension.
  
\begin{figure}[htbp!]
\centering
{\includegraphics [width=0.32\textwidth, angle=0] {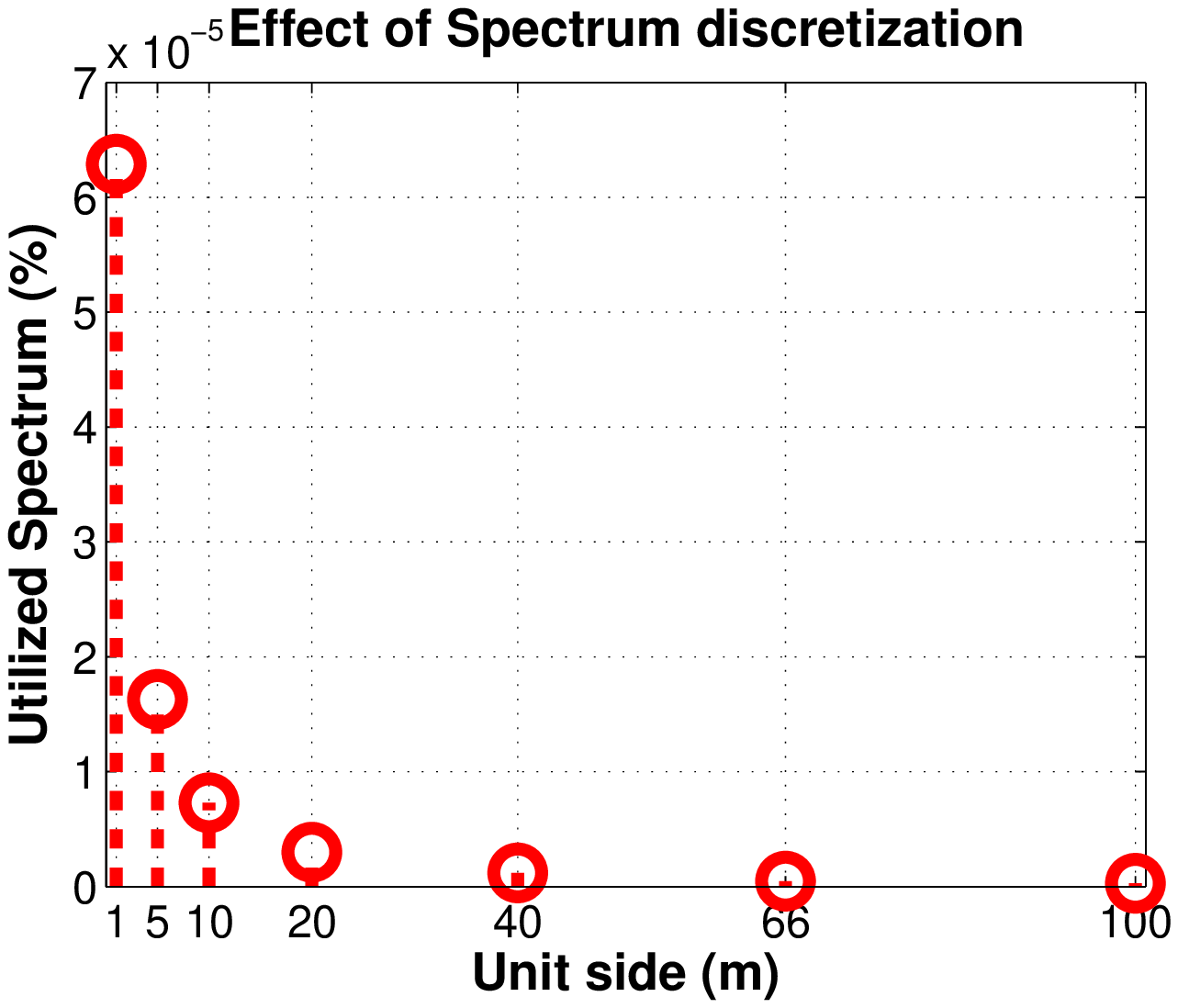}}
{\includegraphics [width=0.32\textwidth, angle=0] {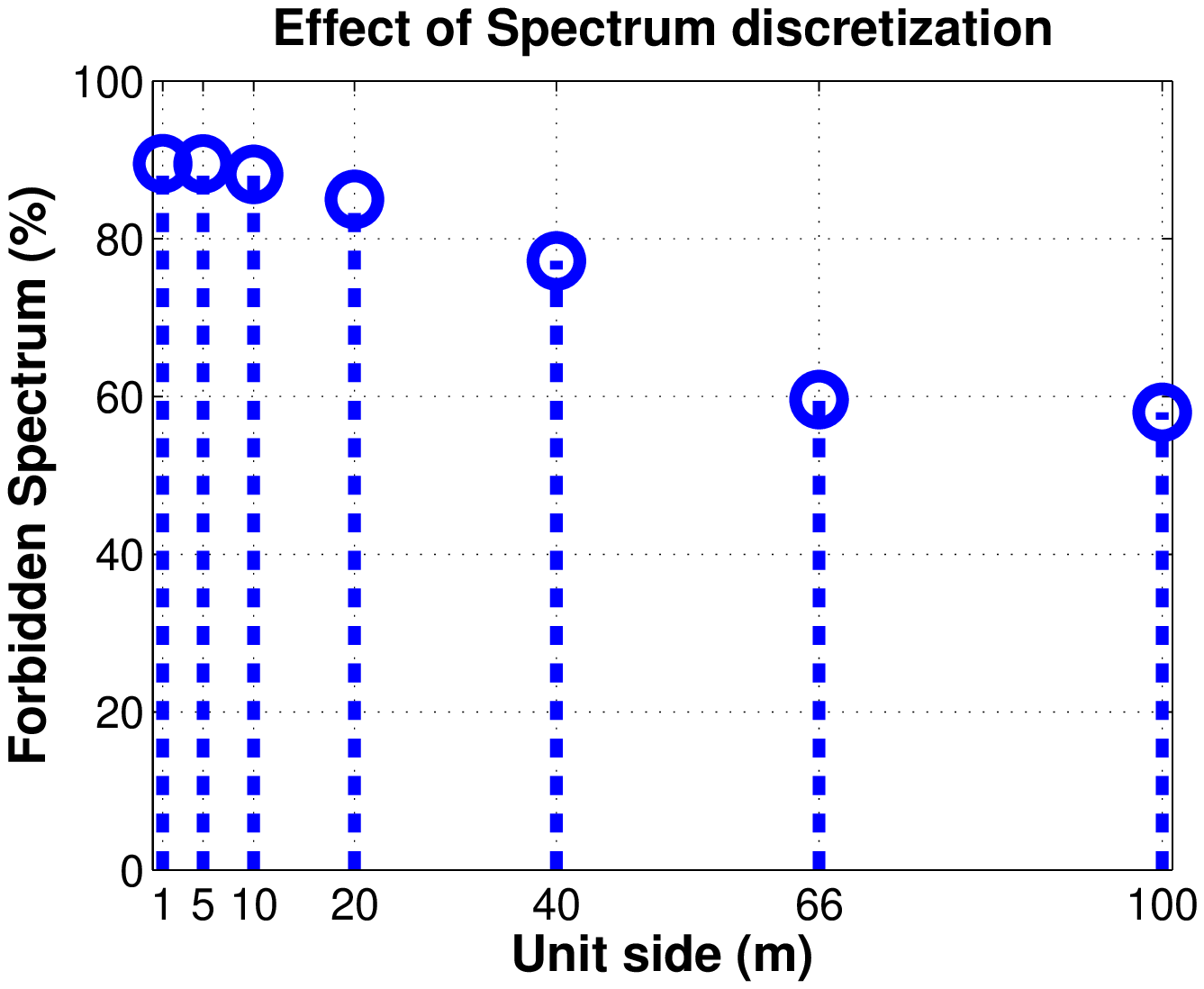}}
{\includegraphics [width=0.32\textwidth, angle=0] {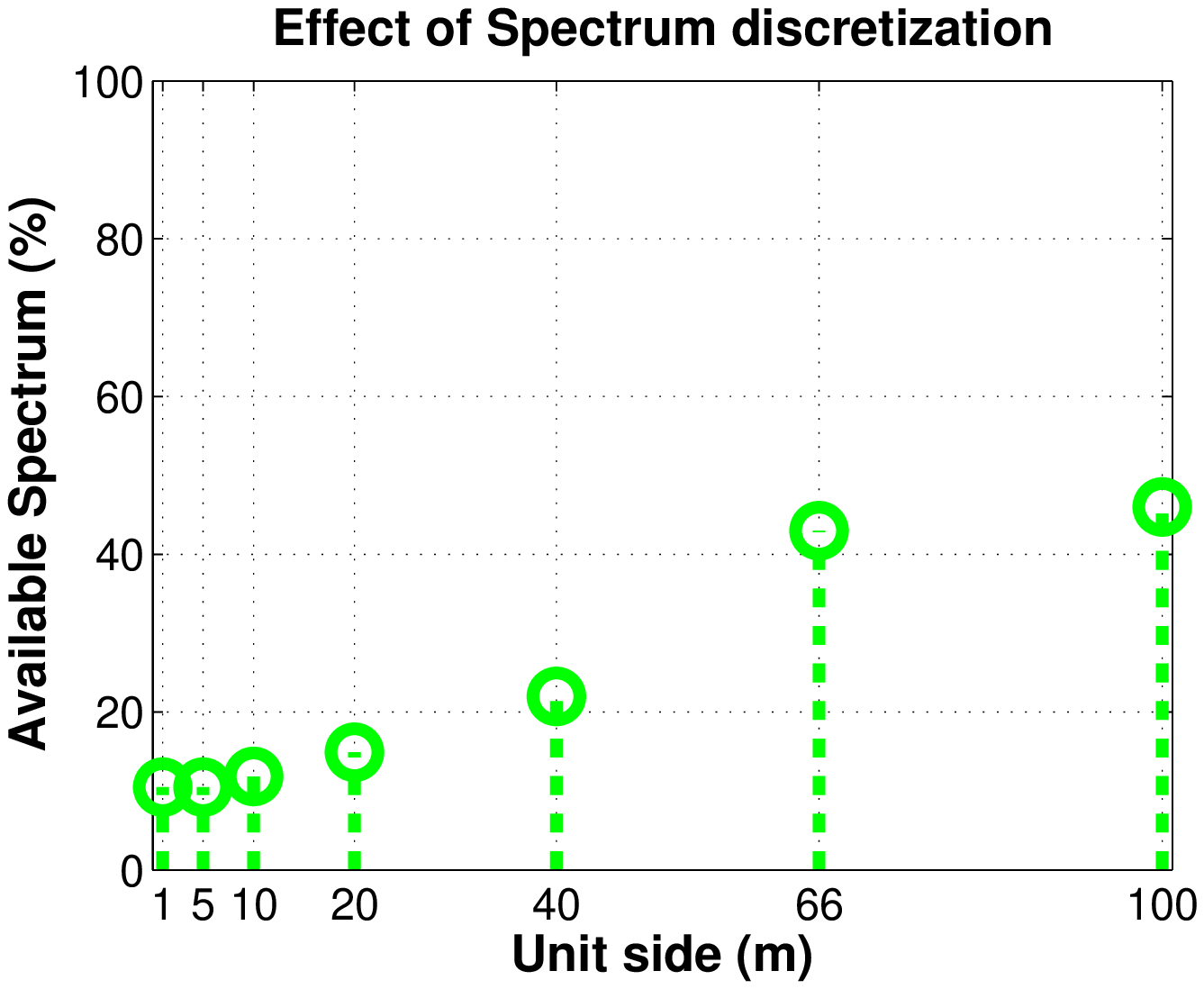}}
\setlength{\abovecaptionskip}{-9pt}
\caption{Spatial sampling of spectrum consumption. The plots capture the effect of spatial sampling rate on the quantification of the spectrum consumption spaces. The X axis shows the length of the hexagonal unit regions. The spectrum consumption in a unit-spectrum-space is governed by spectrum consumption at the \textit{sample point} in the unit-spectrum-space. Thus, we observe that the spectrum consumption spaces are more accurately captured with higher sampling rate.} 
\label{fig:L207}
\end{figure}

Figure \ref{fig:L207} illustrates the impact of spatial sampling rate on quantification of the spectrum consumption spaces. The spatial sampling rate varies from 1 m to 100 m. When the side of the unit-region is 1 m, the quantified value of the total spectrum space is much higher as compared to the total spectrum space when the side of the unit-region is 100 m. We note that the worst-case distance between the sampling point and a transceiver can be half the spatial sampling rate. Here, we use the worst-case setup with the transceiver distances kept maximum from the unit-spectrum-space sampling points. When sampling rate is lower, the spectrum consumed by transmitters and receivers in the unit-spectrum-spaces is more accurately captured by the unit-spectrum-space occupancy and opportunity respectively. Consequently, we observe that the quantity of transceiver consumed spectrum decreases with sampling rate and that of available spectrum increases with sampling rate. 

\subsubsection{Statistical Modeling of Spectrum Consumption}
With MUSE, we choose to capture spectrum consumed by transceivers with a sampling approach instead of using a statistical technique. Thus the methodology captures the \textit{instantaneous use} of the spectrum independent of the characteristics of the propagation environment. It is possible to enrich the representation of the spectrum consumption in a unit-spectrum-space using statistical methods; for example, similar to statistical modeling of the RF-environment, capturing spectrum consumption within a unit-spectrum-space with multiple samples and applying statistical functions \cite{oms4_sce}.
%
\subsubsection{Spectrum Use in the Code Dimension}
MUSE is agnostic of the waveforms employed by the RF-entities. Thus, it does not capture the spectrum use in the \textit{code} dimension of spectrum-access.

\section{Towards Maximizing the Use of Spectrum}
The ability to characterize the use of spectrum provides an insight into the opportunities to maximize the use of spectrum. By articulating the spectrum rights in terms of quantified use of the spectrum, we get the ability to precisely \textit{control} the use of spectrum.

Now, we revisit the question-map from Figure~\ref{fig:qdsa_qneed}. In order to maximize the use of spectrum, we emphasize on  maximizing the spectrum sharing opportunities, maximizing the recovery of the underutilized spectrum and maximizing the exploitation of the recovered spectrum.

\begin{figure}[htbp!]
\begin{center}
{\includegraphics [width=0.84\textwidth, angle=0] {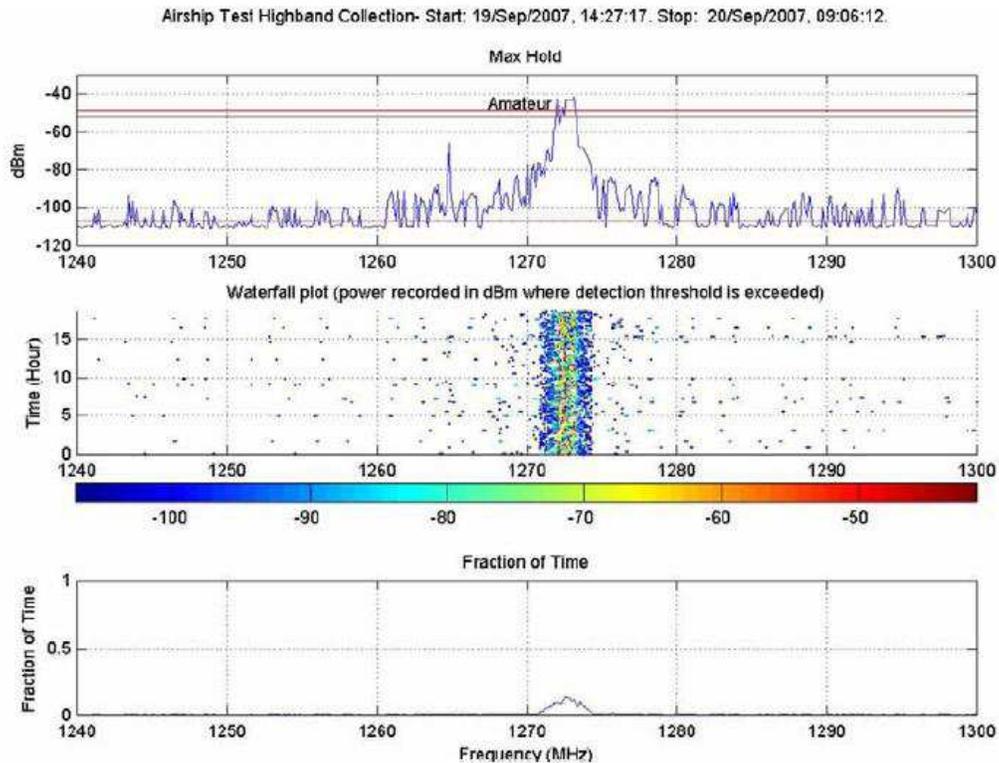}}
\setlength{\abovecaptionskip}{-9pt}
\caption{Spectrum occupancy measurements at Loring Commerce Center, Limestone, Maine during September 18-20, 2007 \cite{ssc_erpek}. The measurements illustrate the fast variations in the spectrum occupancy over time.}
\label{fig:rtla-dsa-ch3-ax13}
\end{center}
\end{figure}

For maximizing the spectrum sharing opportunities, we need to investigate into the spectrum measurements. The spectrum measurements data from \cite{ssc_erpek} illustrates slower to faster degrees of variations in the occupied RF-power. Figure~\ref{fig:rtla-dsa-ch3-ax13} from \cite{ssc_erpek} captures the fast variations in the occupied RF-power over time. For maximizing the use of spectrum, we need to take into account variations in the use of spectrum in the space, time, and frequency dimensions. The case-study of OSA has shown that conservative assumptions based on the worst-case conditions severely limit the spectrum available for sharing \cite{osa_feasib}. Hence, in order to maximize the spectrum available for sharing, we encourage characterizing the use of spectrum in real-time. 

Next, once we observe a significant quantity of the underutilized spectrum within a geographical region, we seek to reuse of the underutilized spectrum using dynamic spectrum sharing paradigm. A \textit{spectrum-sharing policy}\footnote{Here, we distinguish these spectrum-access constraints that imply the spectrum available for sharing from the spectrum-access constraints on an individual RF-entity while exercising a spectrum-access. We call the former one as \textit{spectrum-sharing policy} and the later one as \textit{spectrum-access policy}} under a spectrum sharing model plays a central role in shaping the spectrum available for sharing. The constraints defined in a spectrum-sharing policy may suggest a guard-space in time, frequency, or space dimensions. This implies a certain amount of available spectrum is rendered un-exploitable. In \cite{oms2_sca}, we identified that using the worst-case propagation conditions and using the worst-case receiver positions due to the lack of knowledge of the receiver positions as the key weaknesses of OSA spectrum-sharing policy. Consequently, the minimum sensitivity and the maximum transmit-power constraints imposed on the secondary users tend to be very conservative and the available-spectrum that can be exploited by the secondary users is severely (less than 1\%) limited\cite{oms2_sca}. In \cite{oms3_cf1}, we addressed optimizing the spectrum available for sharing by optimizing the constraints defined under a spectrum-sharing policy while ensuing non-harmful interference to the primary and secondary users.

The actual spectrum exploited by a transceiver depends on the \textit{spectrum-access policy}. In order to dynamically define an optimal spectrum-access policy, we need the ability to recover the spectrum implied available by a spectrum-sharing policy. In \cite{oms4_sce}, we illustrate recovery of the underutilized spectrum using a dedicated external RF-sensor network. In contrast to the traditional approach wherein spectrum holes are inferred by employing detection of the transmitter-signal, this approach estimates unit-spectrum-space opportunity (given by ~\ref{eq:ussop}) across the unit-spectrum-spaces within a geographical region. The RF-sensors characterize the fine-grained propagation environment, estimate the spectrum-access parameters of the transceivers exploiting signal cyclostationarity, and thereby estimate the available spectrum-space. 

In order to optimally exploit the recovered spectrum for satisfying the spectrum-accesses from multiple spatially-overlapping heterogeneous wireless networks, we emphasize the need for defining a quantified spectrum-access policy. A quantified spectrum-access policy identifies spectrum-access rights in terms of how much of the spectrum can be consumed by an RF-entity within each of the unit-spectrum-spaces within a geographical region. Thus, when multiple spatially-overlapping RF-entities are sharing the spectrum, non-harmful interference could be ensured to the receivers in the system. We note that using the RF-sensor network, the dynamically defined spectrum-access rights can be enforced in real-time by estimating the spectrum-consumption spaces for the individual transmitters.

Defining quantified spectrum-access rights requires us to develop spectrum assignment schemes that can \textit{quantitatively} control the spectrum-footprint assigned to each of the transceivers. A quantified approach to spectrum exploitation essentially transforms the spectrum-scheduling and spectrum-allocation problems into a problem of optimizing the spectrum consumption spaces for a set of spectrum-access requests. In \cite{oms3_cf1}, we present maximizing the number of spectrum-access requests based on \textit{spectrum-consumption by an RF-entity}. 

\begin{figure}[t]
\centering
{\includegraphics [width=0.90\textwidth, angle=0] {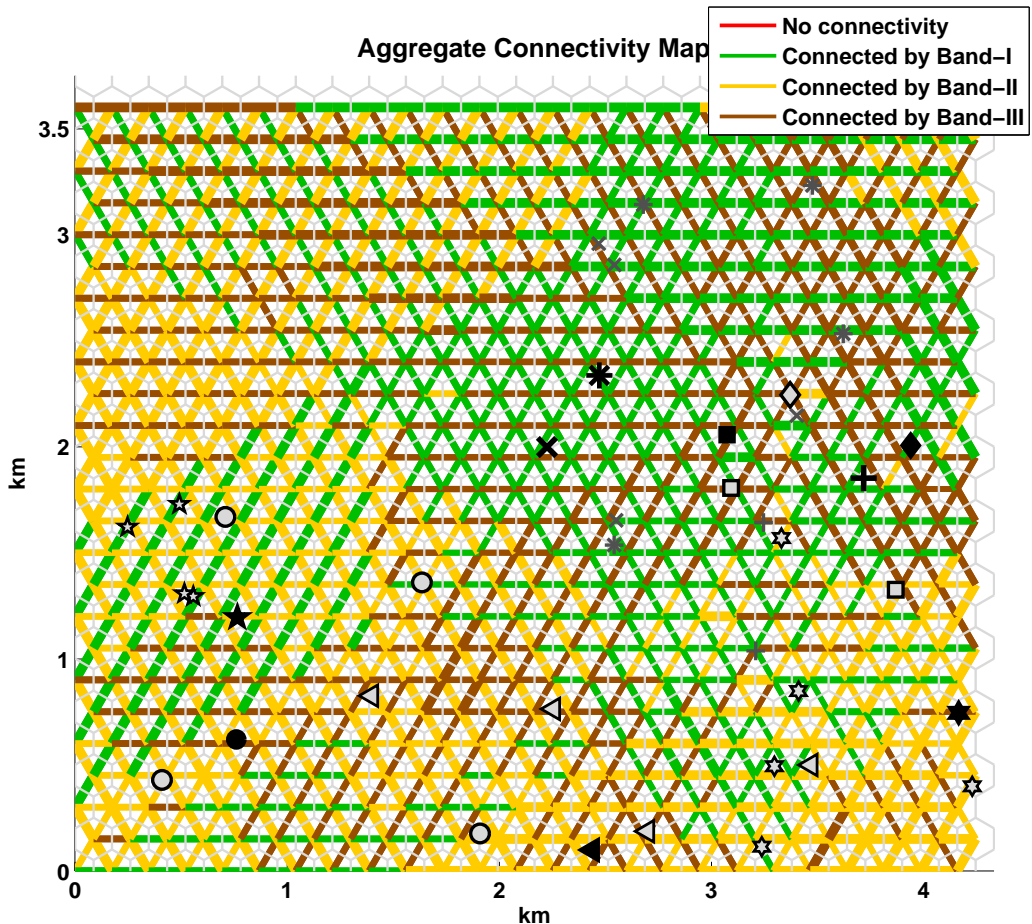}} 
\setlength{\abovecaptionskip}{-12pt}
\caption{Multiple-band RF-connectivity map showing the degree to which adjacent unit spectrum-spaces can connect using a new RF-link. Transmitters and receivers in the same network have the same shape; the transmitter is solid. The particular frequency band is encoded through the color of the connecting lines, and the line color is determined by the best available connectivity. The map reveals exploitable spectrum opportunities in the spatial and frequency dimensions. For this particular set of networks, the spectrum opportunities within the band I (green), band II (yellow), and band III (brown) are easily discerned. Moreover, the directional variation in spectrum opportunity is taken into account, so that the best channel to use depends on the spatial orientation of the to-be-added transmitter-receiver pairs.}
\label{figure:agg_connectivity_map}
\end{figure}

By characterizing the unit-spectrum-space opportunity in the spatial dimension, we can infer spectral connectivity across the adjacent unit-regions within a geographical region using a certain frequency band. We can also combine the spectrum-access opportunities across multiple frequency bands and infer aggregate RF-connectivity map as shown in Figure~\ref{figure:agg_connectivity_map}. Such fine granular characterization of the use of spectrum can be useful in the optimization of scheduling, spectrum assignment, and routing.

We can further improve the amount of spectrum available for sharing by \textit{reducing} the spectrum-consumption by RF-entities. Directional transmission and reception helps to improve SINR; reduce the spectrum consumed by transceivers and increase the available-spectrum \cite{oms3_cf1}. Also, with active role by incumbents, primary transmission power could be increased to enhance SINR at the primary receivers, minimize the receiver consumed spectrum, and maximize the spectrum available for sharing with the secondary services \cite{oms3_cf1}.

The key challenge for optimizing spectrum sharing potential is the \textit{dynamicity of RF-environment}. The propagation conditions may vary quite fast at the order of every few 100 meters. A certain frequency band may not be available at a certain location due to a reappearing primary user. The link quality for secondary access may change due to another mobile secondary user. Here, we note that traditionally wireless network services have been \textit{conservatively handcrafted} to ensure minimum performance under the worst-case conditions. Dynamic spectrum sharing model forces us to come of the constrained setup and develop dynamic responses to the unknown possibilities in terms of the RF environmental and access conditions. Such dynamic response is possible only via learning the RF-environment and synthesizing behavior, decisions, and actions in advance. 

In this regard, the proposed spectrum-discretization approach enables bringing in learning and adaptation to the spectrum management functions. For example, in order to dynamically define the spectrum access rights, we investigate fine granular characterization of the shadowing profile within a geographical region \cite{oms4_sce}. Real-time characterization of the shadowing loss within a fine granular region helps to control the guard space at the finer granularity and maximize the spectrum available for sharing under dynamic RF environments. With regards to adaptation of the spectrum-access, the RF-connectivity map from Figure~\ref{figure:agg_connectivity_map} can be used for provisioning redundancy for RF-links. Thus, analysis of the use of spectrum provides rich information and improves the ability to adapt in case of spectrum mobility events. 
%
\section{Benefits Towards Addressing the Challenges for the Dynamic Spectrum Sharing Paradigm} 
In this section, we enumerate benefits of MUSE from technical, operational, regulatory, and business perspectives. 
\subsection{Benefits Towards Spectrum Management}
\begin{itemize}
  \item MUSE helps to characterize and quantify the use of spectrum at the desired granularity in the space, time, and frequency dimensions. MUSE helps to query how much spectrum is consumed by a single transceiver or any logical collection of the transceivers.   
	\item MUSE helps to compare, analyze, and optimize the performance of spectrum management functions. For example, it is possible to quantitatively analyze performance of ability to recover the underutilized spectrum of various spectrum sensing algorithms (like energy-detection, cyclostationary feature detection) or various cooperative spectrum sensing infrastructures  based on the recovered spectrum space, lost-available spectrum space, and potentially-incursed spectrum space. 
	\item MUSE can help to estimate the available spectrum and the exploited spectrum. Thus, it offers the ability to define the spectrum-access rights based on the real-time RF-environment conditions. Using the real-time RF-environment conditions helps to get rid of conservative assumptions and make an efficient use of the spectrum.
	\item The proposed spectrum-discretization approach facilitates adaptation of the spectrum management functions under dynamic RF environment conditions and dynamic spectrum-access scenarios.  
\end{itemize}

\subsection{Benefits Towards Dynamic Spectrum Access}
\begin{itemize}
  \item MUSE enables us to articulate, define, and enforce spectrum-access rights in terms of the use of spectrum by the individual transceivers.
	\item From operations perspective, the guard space could be effectively controlled. The discretized spectrum management approach enables us to easily map a guard margin value to the amount of the inexercisable spectrum. Thus, depending on the user-scenario, spectrum sharing behavior could be changed with visibility into the implied availability of the spectrum.
	\item Another advantage from an operational perspective is controlling the granularity of spectrum sharing. With discretized approach to spectrum management, the dimensions of a unit-spectrum-space imply the granularity of sharing of the spectrum resource. 
	\item With characterization of spectrum-access opportunity in the space, time, and frequency, MUSE provides the ability to share spectrum without defining a boundary across spectrum uses. 
	\item The discretized spectrum management can be applied independent of the spectrum sharing model. Thus, it can be applied in case of the completely dynamic spectrum sharing model like pure spectrum sharing model or even in case of a conservative spectrum sharing model like static spectrum sharing model.
	\item From a regulatory perspective, MUSE offers the ability to enforce a spectrum-access policy and ensure protection of the spectrum rights of the users. As the spectrum-access rights are identified at the granularity of a single transceiver, the violations by a particular transmitter, or the harmful interference for the individual receivers could be characterized and quantified.
\end{itemize}

\subsection{Benefits Towards Spectrum Trade}
\begin{itemize}
	\item The quantified approach brings in simplicity in spectrum trade. It enables easier understanding and interpretation of the outcomes; thus, it requires less skills of its users. 
	\item The quantified approach enables to investigate the amount of the spectrum that can shared and evaluate the potential for a business opportunity.
	\item From a business development perspective, spectrum sharing models devised using a quantified approach enable spatial overlap of multiple RF-systems and avoid spatial fragmentation of coverage. This is important for defining new services exercising shared spectrum-access rights. 
	\item Aggregation of fine granular spectrum sharing opportunities gives incentives for spectrum-owners to extract more value out of their underutilized spectrum; a bigger spectrum-pool is attractive for secondary users as well. Thus, characterization of the fine granular spectrum-access opportunities enables building a bigger spectrum-resource pool. 
\end{itemize}

Thus, MUSE provides a unified foundation for the spectrum commerce, regulation, operations, and technology.

Finally, a note on the real-time dynamic spectrum access. We encourage defining and enforcing spectrum access rights in real-time. Although this requires a dedicated spectrum management infrastructure, it potentially brings in new business models along with flexible and efficient use of the spectrum and an ability for automated regulation of the dynamic spectrum-accesses. 

\section{Conclusions and Future Research Avenues}
Dynamic spectrum sharing marks a new era in the history of radio communications. With the static and exclusion spectrum allocation paradigm, the spectrum management functions need not explicitly consider the use of spectrum by the receivers. Under the new dynamic spectrum sharing paradigm, multiple spatially-overlapping heterogeneous wireless networks exercise a shared access to the spectrum. This necessitates considering the spectrum used by the individual transmitters \textit{and} receivers.

MUSE provides us the ability to articulate, define, and enforce the spectrum-access rights of the individual transceivers in terms of the spectrum used in the space, time, and frequency dimensions.  This is especially important in terms of addressing the regulatory issues and ensuring protection of the spectrum rights under dynamic spectrum sharing paradigm.

MUSE captures the fine granular variations in the use of spectrum in the space, time, and frequency dimensions. As the demand for spectrum is growing, it is important to exploit these fine granular spectrum-access opportunities and improve the efficiency of the spectrum management functions.

With discretization of the spectrum space, MUSE enables us to quantify the use of spectrum. This ability to quantify the use of spectrum enables to treat spectrum as a commodity. It brings in simplicity, precision, and efficiency into the business models based on the new dynamic spectrum sharing paradigm. From a technical perspective, it facilitates characterizing and quantifying the performance of spectrum management functions directly in terms of the use of spectrum in the space, time, and frequency dimensions. This provides aid to investigate the issues in the recovery and exploitation of the underutilized spectrum. From an operational perspective, the spectrum-space discretization approach provides flexibility in terms of controlling the sharing of spectrum at the desired granularity.

For the past several decades, the static and exclusive spectrum allocation paradigm implied a restricted setup for the spectrum management functions. The new paradigm calls for the ability to learn and adapt under dynamic spectrum-access scenarios and unknown RF-environment conditions. In this regards, using MUSE methodology, we encourage further research in developing deep knowledge of the use of spectrum and the RF-environment conditions in order to synthesize dynamic spectrum management behaviors, decisions, and actions.




\end{document}